# Fortaleza: The Emergence of a Network Hub

Eric Bragion[1], Habiba Akter[2], Mohit Kumar[2], Minxian Xu[3], Ahmed M. Abdelmoniem[1] and Sukhpal Singh Gill[1]

[1]School of Electronic Engineering and Computer Science Queen Mary University of London, UK
[2]Department of Information Technology, National Institute of Technology, Jalandhar, India
[3]Shenzhen Institute of Advanced Technology, Chinese Academy of Sciences, Shenzhen, China

eric.bragion@grupovirta.com.br, h.akter@qmul.ac.uk, kumarmohit@nitj.ac.in, mx.xu@siat.ac.cn, ahmed.sayed@qmul.ac.uk, s.s.gill@qmul.ac.uk

*Abstract*— Digitalisation, accelerated by the pandemic, has brought the opportunity for companies to expand their businesses beyond their geographic location and has considerably affected networks around the world. Cloud services have a better acceptance nowadays, and it is foreseen that this industry will grow exponentially in the following years. With more distributed networks that need to support customers in different locations, the model of one-single server in big financial centres has become outdated and companies tend to look for alternatives that will meet their needs, and this seems to be the case with Fortaleza, in Brazil. With several submarine cables connections available, the city has stood out as a possible hub to different regions, and this is what this paper explores. Making use of real traffic data through looking glasses, we established a latency classification that ranges from exceptionally low to high and analysed 800 latencies from Roubaix, Fortaleza and Sao Paulo to Miami, Mexico City, Frankfurt, Paris, Milan, Prague, Sao Paulo, Santiago, Buenos Aires and Luanda. We found that non-developed countries have a big dependence on the United States to route Internet traffic. Despite this, Fortaleza proves to be an alternative for serving different regions with relatively low latencies.

## I. INTRODUCTION

There were 484 submarine cables connecting all continents, except Antarctica in July 2021 [1], and they are responsible for the transport of 99 per cent of the international data traffic [2]. It is already known that the pandemic has accelerated the digitalisation even in poor countries, and the internet traffic is expected to jump from 2.4 exabytes per day in 2016 to 7.7 exabytes this year, number which is 135 times the figures registered in 2005 [3]. The need for digital content has, consequently, also impacted the data centres industry, more precisely cloud computing services.

The amount spent with cloud infrastructures was higher than on-premises, $130 billion and $90 billion, respectively, in 2020 [4] and the large-scale adoption of cloud solutions that year was due to the needs that COVID-19 imposed [5]. Numbers show that 92 per cent of the companies have a multi-cloud strategy, 82 per cent have public and private clouds and that the majority of enterprises (83 per cent) spend more than $1.2 million per year on cloud solutions, an increase of 11 per cent when comparing to the previous year [5]. However, despite the optimism, there are some points of attention.

The same report identified that cloud costs exceeded budget by, on average, 24 per cent, and that at least 30 per cent of the total cost was considered a waste in 2020 [5]. Because of this, optimise the existing use of cloud, and consequently, save money, leads the list of priorities in 2021 – it is the fifth year in a row in which this topic is listed as the priority number one. Economic groups, such as the European Union and Mercosur, made it easy to trade across markets and being digital also means that more areas can be explored, and new revenue income might be created. China, United States (US), and United Kingdom (UK) are the main countries taking advantage of online transactions, and it is estimated that, this year, almost 20 per cent of the worldwide purchases will be online [6].

Without taking into consideration logistics challenges delivering goods, providing a smooth digital experience is also mandatory to the businesses' success. While in the past networks were mainly centred in global financial centres, now there is a dispersed network that rely more on indirect connections [7]. Cities like Fortaleza, in Brazil, and Marseille, in France, for example, are standing out because of their strategic locations when it comes to global communications and have already established themselves as important indirect connections to financial centres in their regions. The importance of latency has increased over the years and several companies have seen it as a critical factor for the business.

Studies have shown that the bounce probability increases in the same proportion as the page load time on websites: the more it takes, the most likely is the user will give up [8]. For instance, if a website takes more than five seconds to load, 74 % of the users will not continue with the intended task [9]. Moreover, it is estimated that Amazon, which has several business units, including cloud services, has a 2% increase in conversion on its website for every second improved in the speed [9]. Another industry that connectivity has affected directly the business is the e-gaming.

While in the 90s online games were restricted to small groups competing mostly in LANs, it is expected that in four years the game streaming and eSport industry will be worth $3.5 billion, and that Latin America will be a key region in terms of viewers with an estimated audience of 130 million people [10]. With an expected global audience of 1 billion people by 2025 [10], better and reliable networks are required, as well as smoothly communications between servers and users.

### A. Motivation and contributions

As the world gets more and more digital, there is an increased need for low latency to deliver the better user experience [27]. While big organisations have the resources to make use of robust Content Delivery Networks, in general, small enterprises still need to take a close look at their spending and optimise it as much as possible [28].

With countries in different continents speaking the same language, but content production still concentrated in just a few, what was delivered physically before, now needs to be transmitted digitally. As internet is a network of networks and lacks a central organisation that unites information from all the parties involved, it is important to identify alternatives outside big centres and have access to information for better decision-making.

This paper aims to analyse Fortaleza as an international hub connecting Brazil, North America and Africa and evaluate if, overall, the latency among those regions is within acceptable indexes that will result in good user experiences. Nowadays, the city has the largest number of submarine cable connections in the world [11], 16 in total, with routes to North and South America, Europe and Africa, what puts the area at an advantage when it comes to the availability of different routes. Furthermore, the region has been investing massively in network infrastructure: in a public-private initiative, an optical fibre structure of 8,000 kilometres was created in Ceara, stated in which Fortaleza in located, connecting major cities in an attempt to provide high-speed internet access to all public bodies and most of the urban population in the region [12].

Specifically, we want to understand:

- The communication cost between some non-developed countries in those areas;
- The communication cost with among developed countries;
- The communication cost with the most populated city in Latin America.

When identifying and analysing communication cost in terms of latencies, we believe this paper will be useful to different stakeholders when planning infrastructure and content distribution. The main contributions of this project are:

- To facilitate governments to understand the communication costs between different regions and create design policies that will promote better connectivity;
- To demonstrate that different providers might have different service level agreement (SLAs) based on their network and, consequently, varied user experience;
- To show the wholesale companies for the importance of looking glasses for their clients.

*B. Article Organisation*

The first section of this paper introduced the reader to the topic, talked about motivation, contributions and now, the organisation of this research. The remaining content is organised as follows: Section II: findings from previous works are presented, as there are important facts that were taking into consideration in this study. It is interesting to notice that the diversity of authors referred located in North America, Europe and Asia portraits, once again, that latency is a global concern. Section III: addresses the methodology applied to this study, from the creation of classification groups and formats to tools used and data analysed. Section IV: the experimental results are presented by geographic region. Section V: concludes this paper and presents possibilities for future work.

II. RELATED WORK

Previous studies have expressed how latency affects different areas of businesses, from online live events with musicians in different locations to e-commerce and financial transactions [13]. Most recently, with the increased number of games servers around the world, the number of players has also increased considerably, but the communication between servers and user's machines are still pointed as one of the biggest challenges, having latency playing a big role [14][15].

To prove its impact, latency has been simulated modifying games' source code and emulating network issues with Netem, available in some of Linux's distribution [14][15], for example. Some techniques were proposed to improve latency, such as equalizing the routing architecture [13] and redundancy as an alternative, but the latter might increase the overall use of the network [16]. There is a consent that latency affects and influences the user experience [17][18], and it is also clear that users are affected differently based on the actions they are taking [14][17]. For instance, it is known that delays are much more acceptable when watching videos rather than playing games[19].

Some have argued that technological advancements haven't collaborated to reduce network latency as regions nearby still register high latency [20], but there is more beyond technological aspects that affect the network performance, such as political influence and agreements between different networks [21]. Although looking glasses have been highlighted as an important tool to measure network indexes like connectivity and routes [22], it has also been stressed that queuing delays cannot be foreseen precisely yet [20]. HostDime's looking glass tool enables the observation of backbone traffic and network efficacy as it emerges via remote networks [25]. The emergence of new connection areas and the dominance of some nations were also highlighted in the academic world [26] [27] [28].

It is estimated that a considerable amount of the Internet traffic passes through the US, position that will probably remain unchanged for years yet due to the lack of agreement between different networks and the country's geographic location [7]. Despite this, when it comes to network and connections, the dependency on big cities has decreased along the years, although they are still considered important, and indirect connections through other centres are increasing [7].

Laboratory studies have been known as a good approach to simulate real-life scenarios and identify solutions for issues [29]. However, it is a controlled environment, and it might lack unforeseen circumstances that might affect networks [26]. Field network testing, as it is used in this study, gives the opportunity to make use of an existing infrastructure that might also be used by enterprises to deliver their services and/or products [31]. Thus, the likelihood the experiment results are closer to real-life scenarios is higher [30].

When comparing different services/applications and networks, it is important to establish an interchangeably standard measurement that can be used in all models of sampling, regardless of what you have at the application level, such as past works defined levels of online game players and how they are affected by latency [14]. Despite this, none of the studies found determined latency ranges and their level of acceptance for general use, and we have addressed this in this paper. This has also allowed us to identify possible bottlenecks and suggest scenarios that would be more valuable for the study case.

Another point of interest is having the US as a central hub of the internet. Although previous studies have mapped the number of backbones in different regions [7] and this might be an assumption when we see the distribution of submarine cables around the world, for example, it is important to analyse some packets routes to have clear evidence of this.

III. METHODOLOGY

Currently, in general, the market hasn't established what low, medium and high latency should be, and this changes considerably when the object of study is gaming. For the effect of comparison, this paper uses the range shown in TABLE I. , which is based on a survey by a well-known technology company [23]. Ideally, for a better user experience, the latency should be in the exceptionally low, low or average range.

To analyse the effectiveness of one data centre covering more than one geographic region, more than considering it is location, we need to measure the latency between different starting points across the globe to the same destinations, as this gives us a figure of the time response for any command between them. As previous works have stated that latency might affect users differently [14][17], this study focus on four different regions (North and South America, Europe and Africa) that are the target areas of a start-up called Latudio - a language learning app that will be available in seven languages-, which might be considered as a parameter for other use cases.

Currently, the company has a single server in Roubaix, France – AMD Ryzen 7 3700 PRO - 8c/16t - 3.6 GHz/4.4 GHz, 64 GB ECC 2666 MHz, 2×960 GB SSD NVMe, 1 Gbps outgoing bandwidth, 10 Gbps incoming bandwidth – and is experiencing relatively high latency across all regions but Europe as shown in TABLE II. . From 10 cities, six of them have high latency when communicating with the server in France, which shows the company needs to have at least another server covering a different geographic region.

Our simulation utilizes realistic data making use of looking glasses made available by telecommunication providers and uses the ping command to estimate the latency between the locations. To increase the level of accuracy when recording measured latencies, since it is not possible to foresee queues delays [20], we have repeated the test 10 times during different periods. Individually, a ping command sends 4-5 packets to the destination IP and the average of those packets is what we have computed.

In order to compare current latency (Roubaix, France) with a possible more favourable location, for this study, we have chosen Fortaleza, in Brazil, to run the tests as one of the source locations. We also compared the numbers gathered in Fortaleza with the latency registered in Sao Paulo, where there is the largest number of people in the country and data traffic [24]. This will help us to understand the likeness of having indirect connections to big centres when a balance among low latency, content availability across different continents and regions and cost-effectiveness is needed.

In total, including the current provider in France, the network of six companies were analysed (HostIDC[1], Aloo Telecom[2], FDC[3], Globenet[4], OVH[5] and Hostdime[6]), which resulted in 800 latencies registered. Except for the server in Roubaix that is currently used by Latudio and considered as a parameter in this study, all the other organisations were randomly selected based on two criteria: the need of having a public looking glass tool and presence in the cities researched.

TABLE I. LATENCY RANGE AND CLASSIFICATION

| Latency | Classification |
| --- | --- |
| < 20 ms | Exceptionally low |
| 21 to 49 ms | Low |
| 50 to 100 ms | Average |
| > 100 ms | High |

One known IP in each one of the cities studied – Miami (US), Mexico City (Mexico), Frankfurt (Germany), Paris (France), Milan (Italy), Prague (Czech Republic), Sao Paulo (Brazil), Santiago (Chile) Buenos Aires (Argentina) and Luanda (Angola) – was used as the destination for the tests and, except where stated, the latency is the average registered in millisecond (ms).

Besides the average latency, among the 10 latencies registered for each city in each one of the networks, we identified the highest and lowest numbers and computed the difference between them, resulting in the average latency variation. In the cities where the latency variation surpassed 200 milliseconds, we analysed the networks individually to identify the existence of any discrepancy. Within this group, when the lowest and highest latency registered had a significant difference, we ran the *routetrace* command to identify what might be causing such discrepancy.

With the IPs through which the packets passed on the route, we used a "Where is My IP Location" tool[7] that consolidates location information from five different sources, which allows us to have a more precise evidence to determine the device location and, thus, understand if the route taken had any impact on the latency.

IV. EXPERIMENTAL RESULTS

A. Communication with Europe

Communication across nations in Europe has the best indexes in this study when it comes to both latency and latency variation, and results in TABLE II. show that the data centre in Roubaix servers well the entire region. On average, latency is within the exceptionally low and low ranges, between 4 and 24 milliseconds (TABLE III. and Fig. 1), with almost no latency variation (TABLE IV. and Fig. 2) - Milan had 1 millisecond variation, which we can consider as insignificant for any impact on the user experience.

---

[1] *Informações da Rede* (2021). [Tool] Available at: https://lg.hostidc.com.br/. (Accessed: 11 August).
[2] *Looking Glass Aloo* (2021). [Tool] Available at: http://lg.alootelecom.com.br/. (Accessed: 11 August).
[3] *Looking Glass* (2021). [Tool] Available at: https://www.fdcservers.net/looking-glass. (Accessed: 11 August).
[4] *IPv4 and IPv6 Looking Glass* (2021). [Tool] Available at: http://lg.globenet.net/lg/lg.cgi. (Accessed: 11 August).
[5] *OVHcloud Looking Glass* (2021). [Tool] Available at: https://lg.ovh.net/. (Accessed: 11 August).
[6] *Looking Glass* (2021). [Tool] Available at: https://www.hostdime.com/tools/looking-glass/. (Accessed: 11 August).
[7] *Where is My IP Location?* (2021). [Tool] Available at: https://www.iplocation.net/. (Accessed: 11 August).

On the other hand, any server located in South America is not a good option to provide services to Europe, as the latency to all countries is considered high, ranging from 164 to 213 milliseconds. If it was the case to choose between Fortaleza and Sao Paulo, Fortaleza has the lowest latency to all destinations in Europe. However, this route has also the highest latency variation registered (7 to 15 milliseconds) when compared to the two others. Thus, although Sao Paulo to Europe has the highest latency to all destinations when compared to Fortaleza, this route has a better latency variation (2 to 5 milliseconds) than Fortaleza.

Another interesting fact is that, even though Frankfurt is the third furthest city from Fortaleza and Sao Paulo (TABLE V. ), the German city had the lowest latency registered among all the European cities when the source was one of the two cities, which might indicate that better routes/agreements are available.

### B. Communication with North America

When it comes to across countries communication, Miami has the lowest latency from all cities studied (Fortaleza, Roubaix and Sao Paulo), which might support the theory that the United States is a worldwide internet hub. Having the shortest distance to Miami as TABLE V. shows, Fortaleza has also the lowest latency with 79 milliseconds. Even though Roubaix is 499 miles further away from Miami than Sao Paulo, both latencies are almost the same (113 and 114 milliseconds, consecutively), which shows a better connection between Europe and the United States. This is also seen in the latency variation: Roubaix to Miami is the only between continents route that had zero latency variation. On the other side of this measurement, even though Fortaleza to Miami has the lowest latency, it has also the biggest latency variation (TABLE IV. and Fig. 2).

Despite being the neighbours of the United States, Mexico does not take full advantage of being close to an international hub when it comes to networks. The latency from all three cities (Fortaleza, Roubaix and Sao Paulo) to Mexico City was in the high range (above 150 milliseconds) and latency variation was between 27 and 67 milliseconds, with the countries in South America with the most significant variation (TABLE IV. and Fig. 2).

### C. Communication with South America

Communication with Santiago and Buenos Aires has the highest latency across all cities studied, and also the highest latency variation – a difference of up to 642 milliseconds between the lowest and highest latencies (TABLE IV. and Fig. 2).

Although Sao Paulo, Santiago and Buenos Aires are on the same continent and have the shortest distance one from another as shown in TABLE V. , the communication between Roubaix, in France, and Santiago has a lower latency than Sao Paulo and Fortaleza. The same trend is seen on the latency variation: while from Roubaix to Santiago there is a variation of 255 milliseconds, Sao Paulo and Fortaleza have more than twice this figure (TABLE IV. and Fig. 2).

Despite the poor performance in communication with Santiago, the South American cities performed better when the destination was Buenos Aires: Sao Paulo had the lowest latency with 253 milliseconds, followed by Fortaleza (285 milliseconds) and Roubaix (308 milliseconds). Anyhow, all the latency measured are in the high latency range, which means the user might somehow be impact by the low performance.

Among the three cities used as destinations (Sao Paulo, Santiago and Buenos Aires), Sao Paulo has the lowest latency from both Fortaleza and Roubaix. While the latency from Roubaix is in the high latency range, the one from Fortaleza is within the low classification group, which might indicate the city is an alternative to server more than one geographic region. Sao Paulo as a destination has also performed well when measured the latency variation, with the highest difference of 7 milliseconds. Curiously, the latency variation is lower between Roubaix and Sao Paulo than from Fortaleza, which registered a variation 4,5 times greater than the one from Europe (TABLE IV. and Fig. 2). The latency variation between Roubaix and Sao Paulo was almost the same as the variation computed in Sao Paulo, with 1 millisecond of difference.

TABLE II. AVERAGE LATENCY BETWEEN LATUDIO'S SERVER AND CITIES ACROSS THE AMERICAS, EUROPE AND AFRICA

|  | *Latency (ms: milli seconds)* |
|---|---|
| Miami, US | 113 |
| Mexico City, Mexico | 152 |
| Frankfurt, Germany | 8 |
| Paris, France | 4 |
| Milan, Italy | 16 |
| Prague, Czech Republic | 24 |
| Sao Paulo, Brazil | 194 |
| Santiago, Chile | 255 |
| Buenos Aires, Argentina | 308 |
| Luanda, Angola | 197 |

To better understand why the latency and latency variation to Santiago and Buenos Aires are so high, we ran the *traceroute* command from the cities with the highest average latency variation registered to know the route the packets passed through.

Fig. 3 shows the traceroute from Sao Paulo to Buenos Aires. When we analysed the geolocation of the IPs, we identified that the packets went from Sao Paulo to the United States, where they travelled around some cities to go back to Sao Paulo again. Just after this journey, the packets were then sent to Buenos Aires, in Argentina. Clearly, the high latency is due to the journey to the United States to go back to the same location as from where the packets were originally sent. When analysing the networks performance individually, one of them stand out for having an average latency to Buenos Aires that is less than half the general average for Sao Paulo (117 milliseconds) and the lowest latency of 35 milliseconds, which shows us that, if needed, companies have a good route available between the two cities.

The route Fortaleza to Santiago has also showed dependency of the United States, as Fig. 4 shows. The packets go from Fortaleza to the United States, where they also pass through several points of connection, and then go straight to Santiago, without passing through Brazil this time.

## D. Communication with Africa

The communication with Luanda is the only one that both Fortaleza and Sao Paulo have better latency variation than Roubaix. When comparing Fortaleza and Roubaix as the sources, the latter has a latency variation 12 times greater than the prior (TABLE III. and Fig. 1). Such variation from Fortaleza was noticed just within the country (Fortaleza to Sao Paulo) and in the communication with developed nations – all latencies variations with non-developed countries were greater than 200 milliseconds. Despite the good performance in latency variation, the latency to Luanda from the three source cities are within the high latency range. On average, Fortaleza had the best performance with 150 milliseconds, which is almost the same latency to Mexico City, followed by Roubaix (197 milliseconds) and Sao Paulo (226 milliseconds). If we compare just the average latency of this study, we can say that Luanda is better connected to other nations than Santiago and Buenos Aires.

TABLE III. AVERAGE LATENCY BETWEEN CITIES IN MILLISECONDS (MS)

|  | From |  |  |
| --- | --- | --- | --- |
| To | Fortaleza | Roubaix | Sao Paulo |
| Miami | 79 | 113 | 114 |
| Mexico City | 152 | 152 | 185 |
| Frankfurt | 164 | 8 | 189 |
| Paris | 167 | 4 | 195 |
| Milan | 191 | 16 | 213 |
| Prague | 177 | 24 | 203 |
| Sao Paulo | 49 | 194 | 1 |
| Santiago | 298 | 255 | 370 |
| Buenos Aires | 285 | 308 | 253 |
| Luanda | 150 | 197 | 226 |

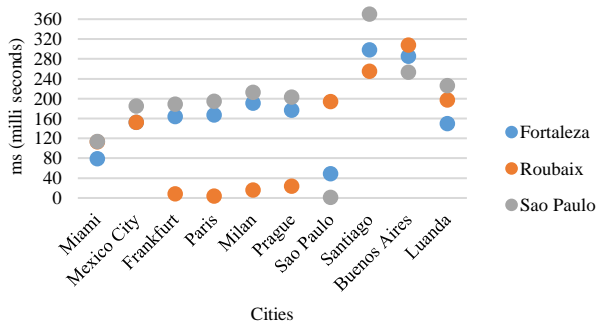

Fig. 1. Average latency between cities in milliseconds (ms)

When each network was analysed individually, there were two data centres in Fortaleza that had latencies to Luanda that were around half of the average. To better understand this, we ran the traceroute command in one network that had a low latency and another one with high latency, as Fig. 5 and Fig. 6 show. As with Santiago and Buenos Aires, high latency is experienced when the packets are sent to North America. As Fig. 5 shows, the data travels from Fortaleza to the United States, where it goes around some cities, to then finally be routed to Luanda. On the other hand, a direct route between the two countries significantly reduces the latency to almost half the average. Fig. 6 shows that the packets travel from Fortaleza to Luanda without the need to pass by any other city.

## V. CONCLUSIONS AND FUTURE WORK

We measured the latency between different locations having Roubaix, Fortaleza and Sao Paulo as the starting points and Miami, Mexico City, Frankfurt, Paris, Milan, Prague, Sao Paulo, Santiago, Buenos Aires and Luanda as the destinations to analyse if Fortaleza can be considered a hub connecting Brazil, North America and Africa. We also wanted to understand the latency between some non-developed, developed countries and specific cities.

TABLE IV. AVERAGE LATENCY VARIATION BETWEEN CITIES IN MILLISECONDS (MS)

|  | From |  |  |
| --- | --- | --- | --- |
| To | Fortaleza | Roubaix | Sao Paulo |
| Miami | 8 | 0 | 6 |
| Mexico City | 67 | 27 | 66 |
| Frankfurt | 15 | 0 | 4 |
| Paris | 9 | 0 | 4 |
| Milan | 7 | 1 | 5 |
| Prague | 15 | 0 | 2 |
| Sao Paulo | 9 | 2 | 1 |
| Santiago | 642 | 255 | 604 |
| Buenos Aires | 242 | 325 | 372 |
| Luanda | 21 | 256 | 42 |

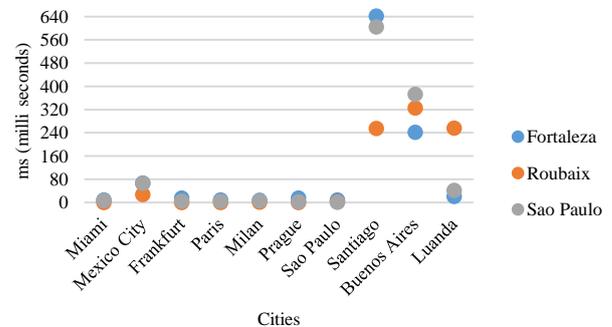

Fig. 2. Average latency variation between cities in milliseconds (ms)

TABLE V. DISTANCE BETWEEN CITIES IN A STRAIGHT LINE IN MILES

|  | From |  |  |
| --- | --- | --- | --- |
| To | Fortaleza | Roubaix | Sao Paulo |
| Miami | 3451 | 4578 | 4079 |
| Miami | 3451 | 4578 | 4079 |
| Mexico City | 4411 | 5701 | 4616 |
| Frankfurt | 4654 | 245 | 6107 |
| Paris | 4381 | 131 | 5841 |
| Milan | 4482 | 455 | 5915 |
| Prague | 4862 | 496 | 6304 |
| Sao Paulo | 1472 | 5958 | - |

|  | *From* | | |
|---|---|---|---|
| *To* | *Fortaleza* | *Roubaix* | *Sao Paulo* |
| Santiago | 2914 | 7339 | 1605 |
| Buenos Aires | 2485 | 6980 | 1042 |
| Luanda | 3570 | 4155 | 4072 |

a. Source: How Far is it Between (2021). [Tool] Available at: https://www.freemaptools.com/how-far-is-it-between.htm. (Accessed: 11 August).

```
traceroute to 170.78.75.88 (170.78.75.88), 30 hops max, 52 byte packets
1   200.16.69.122 (200.16.69.122) 6.124 ms 7.484 ms 6.094 ms
2   200.16.69.44 (200.16.69.44) 43.905 ms 41.138 ms 39.082 ms
3   200.16.69.40 (200.16.69.40) 112.661 ms 112.916 ms 112.788 ms
4   63.217.112.201 (63.217.112.201) 127.551 ms 127.691 ms 127.545 ms
5   63.223.40.2 (63.223.40.2) [AS 3491] 130.175 ms 127.902 ms 127.818 ms
6   63.223.40.2 (63.223.40.2) [AS 3491] 129.443 ms 129.892 ms 146.049 ms
7   94.142.107.30 (94.142.107.30) [AS 12956] 125.888 ms 127.646 ms 128.042 ms
8   94.142.117.59 (94.142.117.59) [AS 12956] 152.260 ms * 94.142.117.83 (94.142.117.83) [AS 12956] 151.991 ms
    MPLS Label=64850 CoS=1 TTL=1 S=1
9   213.140.39.119 (213.140.39.119) [AS 12956] 265.889 ms * 5.53.7.238 (S.53.7.238) [AS 12956] 277.127 ms
10  213.140.39.119 (213.140.39.119) [AS 12956] 290.232 ms 285.719 ms *
11  209.13.168.146 (209.13.168.146) [AS 10834] 279.698 ms 213.140.39.119 (213.140.39.119) [AS 12956] 271.403 ms 209.13.168.146 (209.13.168.146) [AS 10834] 284.447 ms
12  200.32.34.186 (200.32.34.186) [AS 10834] 284.431 ms 288.447 ms 282.176 ms
13  200.16.206.6 (200.16.206.6) [AS 10834] 270.837 ms 277.354 ms 283.317 ms
```

Fig. 3.  Traceroute Sao Paulo to Buenos Aires going through the US

```
traceroute to 192.140.56.150 (192.140.56.150), 30 hops max, 60 byte packets

1   190.15.105.29 (190.15.105.29) 16.850 ms
2   *
3   *
4   64.191.232.76 (64.191.232.76) 67.770 ms
5   100ge8-1.corel.nyc4.he.net (184.104.195.21) 146.019 ms
6   100ge16-1.core1.ashi.he.net (184.105.223.165) 157.361 ms
7   100ge13-1.corel.atli.he.net (184.105.80.162) 145.109 ms
8   100ge0-35.core2.jax1.he.net (72.52.92.50) 145.908 ms
9   100ge3-2.corel.mial.he.net (72.52.92.49) 145.604 ms
10  cl-phei-as263237.e0-51.switch1.mial.he.net (216.66.61.2) 213.081 ms
11  *
12  cli.enduserexp.com (192.140.56.150) 208.924 ms !X
```

Fig. 4.  Traceroute Fortaleza to Santiago going through the US

```
traceroute to 41.223.158.4 (41.223.158.4), 30 hops max, 60 byte packets
wed
1   ***
2   45.238.96.222 (45.238.96.222) 0.307 ms 0.288 ms 0.236 ms
3   200.16.69.2 (200.16.69.2) 66.138 ms 82.118 ms 82.122 ms
4   eqix-dc2.angolacables.com (206.126.238.56) 81.490 ms 81.492 ms 81.924 ms
5   ***
6   102.130.68.194 (102.130.68.194) 303.146 ms 304.813 ms 304.227 ms
7   102.130.68.110 (102.130.68.110) 304.020 ms 305.306 ms 305.743 ms
8   197.149.151.46 (197.149.151.46) 305.745 ms 306.492 ms 307.248 ms
9   41.74.240.97 (41.74.240.97) 306.237 ms 305.255 ms 306.236 ms
10  * 192.168.162.26 (192.168.162.26) 310.650 ms 310.609 ms
11  192.168.13.58 (192.168.13.58) 314.414 ms 315.279 ms 314.861 ms
```

Fig. 5.  Traceroute Fortaleza to Luanda going through the US

```
traceroute to 41.223.158.4 (41.223.158.4), 30 hops max, 60 byte packets

1   190.15.105.29 (190.15.105.29) 16.865 ms
2   48.235.238.170.angolacables.ao (170.238.235.48) 17.124 ms
3   170.238.232.153 (170.238.232.153) 78.490 ms
4   170.238.232.146 (170.238.232.146) 78.245 ms
5   102.130.68.105 (102.130.68.105) 79.255 ms
6   197.149.151.46 (197.149.151.46) 80.629 ms
7   41.74.240.97 (41.74.240.97) 80.285 ms
```

Fig. 6.  Traceroute Fortaleza to Luanda without going through the US

It was clear that any server in South America is not a good option in terms of latency when communicating with Europe, and that the server in Roubaix has an exceptionally low or low latency to any of the European countries studied. Although the latency between Roubaix and Miami are within the high latency range, it might still be considered for businesses purposes, as there is no latency variation, and it had the second-best latency. The communication with non-developed countries showed to be challenging mainly because of the dependency on the United States to route packets to the final destinations. This has highly affected not just the latency, but also the latency variation, having been registered over 600 milliseconds of variation between the lowest and highest average computed.

Some routes, such as Fortaleza to Luanda and Sao Paulo to Buenos Aires, stood out due to some specific networks have considerably lower latency than the average mainly because of more direct connections between the two points [26]. Finally, when it comes to cost efficiency (one data centre used for different regions), content availability in different geographic location and relatively low latency, Fortaleza has showed to be a good option to server big centres in Brazil, the United States, through Miami, and Africa, through Luanda with the exception that the network to be used must be previously analysed to verify its effectiveness. In 1980, a line proposed by Willy Brandt divided the world in two: rich nations in the North hemisphere and poor countries in the South. When transposing the Brandt Line to the distribution and efficiency of IP networks, similar results are found:

- Countries in the North are better connected;
- Communication between Northern nations have almost no latency variation in most cases;
- Connections in the South are more unstable;
- Latency is higher in the South.

Further studies are planned to analyse why the reason why the connection agreements between non-development nations are so poor and/or non-existent, and how the dependency on the United States might affect global communications if something goes wrong. Moreover, the analysis of the connections between Fortaleza and other destinations might indicate other routes that data centres in the city can cover satisfactorily.

DECLARATION OF COMPETING INTEREST

The authors declare that they have no known competing financial interests or personal relationships that could have appeared to influence the work reported in this paper.

ACKNOWLEDGMENTS

We wish to thank Latudio's co-founders, Mark Shimada and Vitek Rozkovec, HostDime's data centre manager, Lucas Montarroios, the managing director at Interxion France, Fabrice Coquio, and the product coordinator at Angola Cables, Edivan Silva, for their availability and insights pre-research regarding submarine networks and global communication. This work is partially funded by Chinese Academy of Sciences President's International Fellowship Initiative (Grant No. 2023VTC0006), National Natural Science Foundation of China (No. 62102408), Shenzhen Industrial Application Projects of undertaking the National key R & D Program of China

(No.CJGJZD20210408091600002) and Shenzhen Science and Technology Program (Grant No. RCBS20210609104609044). We also declare that this work has been submitted as an MSc project dissertation in partial fulfilment of the requirements for the award of degree of Master of Science submitted in School of Electronic Engineering and Computer Science of Queen Mary University of London, UK is an authentic record of research work carried out by Eric Bragion (first author) under the supervision of Sukhpal Singh Gill (last author) and refers other researcher's work which are duly listed in the reference section. This MSc project dissertation has been checked using Turnitin at Queen Mary University of London, UK and submitted dissertation has been stored in 394 repository for university record.